\documentstyle[prb,eqsecnum,preprint,aps]{revtex}
\begin{document}
\draft

\title{A Novel 2D Folding Technique for Enhancing Fermi Surface \\
       Signatures in the Momentum Density: Application to  \\
       Compton Scattering Data from an Al-3at\%Li Disordered Alloy
}
\author{I. Matsumoto}
\address{
         Department of Synchrotron Radiation Science, The Graduate University
         for Advanced Studies, \\
         Tsukuba, Ibaraki 305-0801, Japan
}
\author{J. Kwiatkowska and F. Maniawski}
\address{
         H. Niewodniczanski Institute of Nuclear Physics, Radzikowskiego 152,
         31-342 Cracow, Poland
}

\author{M. Itou}
\address{
         Spring-8, Kamigouri, Akougun, Hyogo 679-5198, Japan
}
\author{H. Kawata}
\address{
         Photon Factory, High Energy Accelerator Research Organization,
         Tsukuba, Ibaraki 305-0801, Japan
}
\author{N. Shiotani}
\address{
         Tokyo University of Fisheries, Kounan, Minato, Tokyo 108-8477, Japan
}
\author{S. Kaprzyk}
\address{
         Department of Physics, Northeastern University, Boston, MA 02115,
USA \\
and  \\
         Academy of Mining and Metallurgy, Al. Mickiewicza 30, Cracow, Poland
}
\author{P.E. Mijnarends}
\address{
         Department of Physics, Northeastern University, Boston, MA 02115,
USA \\
and  \\
 Interfaculty Reactor Institute, Delft University of Technology, Delft,
The Netherlands
}
\author{B. Barbiellini and A. Bansil}
\address{
         Department of Physics, Northeastern University, Boston, MA 02115, USA
}
\date{\today}
\maketitle

%%%

\begin{abstract}

We present a novel technique for enhancing Fermi surface (FS)
signatures in the 2D distribution obtained after the 3D momentum
density in a crystal is projected along a specific direction in
momentum space. These results are useful for investigating
fermiology via high resolution Compton scattering and positron
annihilation spectroscopies. We focus on the particular case of
the (110) projection in an fcc crystal where the standard
approach based on the use of the Lock-Crisp-West (LCW) folding
theorem fails to give a clear FS image due to the strong overlap
with FS images obtained through projection from higher Brillouin
zones. We show how these superposed FS images can be disentangled
by using a selected set of reciprocal lattice vectors in the
folding process. The applicability of our partial folding scheme
is illustrated by considering Compton spectra from an Al-3at\%Li
disordered alloy single crystal. For this purpose, high
resolution Compton profiles along nine directions in the (110)
plane were measured. Corresponding highly accurate theoretical
profiles in Al-3at\%Li were computed within the local density
approximation (LDA)-based Korringa-Kohn-Rostoker coherent
potential approximation (KKR-CPA) first-principles framework. A
good level of overall accord between theory and experiment is
obtained, some expected discrepancies reflecting electron
correlation effects notwithstanding, and the partial folding
scheme is shown to yield a clear FS image in the (110) plane in
\linebreak Al-3\%Li.

\end{abstract}


%%%

\narrowtext
\section{introduction}\label{sec:intro}

Recent advances in synchrotron light sources and detector
technology have rejuvenated interest in high resolution Compton
scattering as a probe of fermiology and electron correlation
effects in wide classes of
materials.\cite{1,2,3,4,5,6,7,8,9,10,11,12,13,14} In a Compton
scattering experiment one measures a directional Compton profile
(CP), $J(p_z)$, which is related to the twice integrated
ground-state electron momentum density $\rho (\bbox{p})$ by
\begin{equation}
\label{eq1}
J(p_z) = \int\!\!\!\int\rho(\bbox{p})dp_x dp_y,
\end{equation}
where $p_z$ is taken along the scattering vector of the X-rays.
In the independent particle approximation,
\begin{equation}
\label{eq2}
\rho (\bbox{p}) = \sum_{b,\bbox{k}}
\left| \int \psi_{b,\bbox{k}}(\bbox{r})
e^{-i \bbox{p} \cdot \bbox{r}} d\bbox{r} \right|^2,
\end{equation}
where $\psi_{b,\bbox{k}} (\bbox{r})$ is the electron wavefunction
in band $b$ and state $\bbox{k}$. The summation in Eq.\ (\ref{eq2})
extends over all initial states that are excited in the scattering
process. The CP thus contains signatures
of the Fermi surface (FS) breaks and correlation effects in the
underlying three-dimensional (3D) momentum density
$\rho(\bbox{p})$.

The determination of FS geometry via the Compton spectroscopy is
complicated by the presence of the double integral in
Eq.\ (\ref{eq1}). As a result, FS breaks in $\rho(\bbox{p})$ do not
in general induce rapid variations in $J(p_z)$, but only in the
first derivative of the profile. Moreover, breaks in
$\rho(\bbox{p})$ caused by FS crossings are scattered throughout
momentum space with weights given by the appropriate matrix
element. Therefore, if this matrix element happens to be small in a
specific case,\cite{14a} then the related FS feature in the CP will
be intrinsically weak and difficult to measure. The standard
approach to tackle these problems is to measure CP's along many
different directions, and invert Eq.\ (\ref{eq1}) to obtain
$\rho(\bbox{p})$ by using one of the available reconstruction
methods.\cite{footnote0,15} The matrix element effects in
$\rho(\bbox{p})$ can then be removed in principle by constructing
the occupation number density in $\bbox{p}$ space, $N(\bbox{p})$,
via the so-called Lock-Crisp-West (LCW) folding procedure\cite{16}:
\begin{equation}\label{eq7}
N(\bbox{p}) \equiv
\sum_{\bbox{G}}\rho (\bbox{p}+\bbox{G})
= \sum_{\bbox{G}} \sum_{b,\bbox{k}} \delta_{\bbox{k},
\bbox{p} +\bbox{G}},
\end{equation}
where the sum over $\bbox{G}$ runs over all reciprocal lattice vectors.
The right hand side of Eq.\ (\ref{eq7}) represents the occupation
number density in the repeated zone scheme.\cite{footnote1}

In practical applications, an accurate reconstruction of
$\rho(\bbox{p})$ along the preceding lines requires the demanding
task of making CP measurements along a suitable group of
directions. However, in many cases the 2D projection of the
momentum density, $\rho (p_x, p_y) = \int \rho(\bbox{p}) dp_z$, and
the corresponding 2D occupation number density, $N(p_x, p_y)$,
along a judiciously chosen direction can give most of the important
features of the FS. As shown in Sec.\ \ref{sec:rec} it is then
sufficient to measure a set of CP's with scattering vectors all
lying in the plane of projection, yielding a significant reduction
of measuring time. The (110) plane in cubic lattices encompasses
many of the high symmetry points and lines in the Brillouin zone
and, therefore, a cross-section in this plane captures much of the
character of the underlying 3D Fermi surface. In this vein, the
(110) projection has been exploited in a number of Compton
scattering and positron annihilation
investigations\cite{17,18,19,20} on {\em bcc} crystals where the
(110) projection turns out to be reasonably free of complications
of overlapping FS images from higher Brillouin zones. In sharp
contrast, it has not been possible to use this approach in {\em
fcc} crystals due to overlap problems which often distort the FS
image severely.

With this motivation, we have examined the question of how the
aforementioned problems of overlap may be avoided in a projected 2D
momentum distribution. Our analysis reveals that the problem can be
solved by using a selected subset of reciprocal vectors in the
folding procedure of Eq.\ (\ref{eq7}). To our knowledge, such an
approach has not been presented previously in the literature.
Although we focus in this article explicitly on the specific case
of the (110) projection in an fcc crystal, these ideas are
generalized straightforwardly to other projections in cubic
lattices. The treatment of projections in general lattices will be
taken up elsewhere.

We illustrate the applicability of our novel folding procedure by 
considering Compton spectra from an Al-3\%Li disordered alloy (fcc) 
single crystal specimen. For this purpose, nine high resolution
CP's have been measured along a set of directions in the (110) plane, 
and corresponding highly accurate theoretical profiles have been computed
within the KKR-CPA first-principles framework. Intercomparisons 
between theory and experiment, and between results based on the use of
a selected and a full set of reciprocal lattice vectors in the folding
process, clearly show the value of our partial folding approach in
enhancing FS signatures in the data. 

Some comment on our choice of the Al-3\%Li disordered alloy for
this study is appropriate. Al, which possesses an almost spherical
FS (in the extended zone), is an interesting base metal for
exploring the evolution of fermiology with changes in the
electrons/atom ratio via alloying on the one hand, and for
investigating electron correlation effects on the momentum density
in multivalent systems on the other. Metallurgically, Al is quite
"exclusive" in that only a few percent impurities can be dissolved
in the solid solution $\alpha$ phase.\cite{21} For example, in
Al-Li the $\alpha$ phase extends at room temperature to only about
3 at\% Li concentration. Al-Li alloys, with the possible addition
of other elements, are also of great technological interest due to
their high strength-to-weight ratio.\cite{22} For these fundamental
and practical reasons, the electronic structure and the nature of
bonding in Al-Li alloys have drawn considerable
interest.\cite{23,24,25} Insofar as the Compton technique is
concerned, relatively little by the way of high resolution work on
disordered alloys is available in the literature despite the fact
that, unlike traditional spectroscopies which rely on quantum
oscillatory phenomena, Compton scattering is not affected by short
electron mean free paths and is thus particularly suited for
investigating disordered alloys. For reasons outlined in this
paragraph, there is a strong motivation for undertaking a
state-of-the-art Compton study of the Al-Li system as a means of
gaining insight into the electronic structure of the system and the
possible role of fermiology in destabilizing the $\alpha$ phase of
Al.\cite{footnote1a}

This article is organized as follows. Section\ \ref{sec:rec}
briefly delineates some reconstruction aspects needed later. It is
shown that it suffices to measure a set of CP's with scattering
vectors all lying in one plane to obtain the 2D projection of the
momentum density onto that plane. In Sec.\ \ref{sec:fold}, the new
folding procedure involving selected reciprocal lattice vectors is
formulated. Section\ \ref{sec:exp} gives pertinent details of the
Compton experiments, while Sec.\ \ref{sec:the} briefly describes
the relevant computational details involved in obtaining
theoretical momentum densities and Compton profiles. Section\
\ref{sec:appl} presents and discusses the new theoretical and
experimental Compton results as well as the applicability of our
partial folding procedure to these data. Some concluding remarks
are collected in Sec.\ \ref{sec:sum}.

%%%

\section{Reconstruction}\label{sec:rec}

The 3D momentum density can be reconstructed by using the
Fourier-Bessel method in which a so-called $B(\bbox{r})$ function
is defined by
\begin{equation}
\label{eq3}
B(\bbox{r}) =
     \int\rho (\bbox{p}) \exp (-i\bbox{p}\cdot \bbox{r}) d\bbox{p}.
\end{equation}
It follows from Eqs.\ (\ref{eq1}) and (\ref{eq3}) that
\begin{equation}
\label{eq4}
B(0,0,z) = \int J(p_z)\exp (-ip_zz)dp_z.
\end{equation}
Thus, from a number of measured CP's, corresponding $B(\bbox{r})$'s
are obtained on radii in $\bbox{r}$ space given by the directions
of the respective scattering vectors. Next, a fine mesh is set up
in $\bbox{r}$ space and $B(\bbox{r})$ is obtained at every mesh
point by interpolation. This makes it then possible to carry out
the inverse Fourier transform
\begin{equation}\label{eqa1}
\rho(\bbox{p}) = (2\pi)^{-3} \int B(\bbox{r})
                    \exp (i\bbox{p}\cdot \bbox{r}) d\bbox{r}
\end{equation}
to obtain $\rho(\bbox{p})$. We refer the reader to
Ref.\ \onlinecite{15} for a detailed discussion of our
reconstruction scheme.

Here we briefly show how a 2D projection of the momentum density
along a given direction can be reconstructed from Compton
measurements in which all the scattering vectors lie in the
perpendicular plane. Without loss of generality, we take [001] to
be the projection direction. From the inverse Fourier transform
Eq. (\ref{eqa1}) the projection onto the (001) plane is:
\begin{eqnarray}\label{eqa2}
  \rho(p_x,p_y) = \int \rho (\bbox{p}) dp_z
           = (2\pi)^{-3} \int dp_z \int \int \int B(x,y,z)  \nonumber \\
   \times \exp[i(p_x x + p_y y + p_z z)] dx dy dz.
\end{eqnarray}
The integration with respect to $p_z$ produces $2\pi \delta(z)$, yielding
\begin{equation}\label{eqa3}
 \rho(p_x,p_y) = (2\pi)^{-2} \int \int B(x,y,0)
                            \exp[i(p_x x + p_y y)] dx dy.
\end{equation}
But, as seen from Eq. (\ref{eq4}) above, $B(x,y,0)$ can be obtained by
interpolation on a fine mesh in the $(x,y)$ plane from measurements in
which all scattering vectors lie in that plane.

%%%

\section{Formulation of Partial Folding Technique}
\label{sec:fold}
This section delineates briefly how the (110) projection of the LCW-folded
momentum density in Eq.\ (\ref{eq7}) can be divided for an fcc crystal
into two identical patterns where one pattern is shifted by $2\pi /a$
along the [001] direction with respect to the other.

We consider an fcc lattice with primitive translation vectors:
\begin{equation}\label{aeq1}
\bbox{a}=\frac{a}{2} (\widehat{\bbox{x}} + \widehat{\bbox{y}})~~;~~
\bbox{b}=\frac{a}{2} (\widehat{\bbox{y}} + \widehat{\bbox{z}})~~;~~
\bbox{c}=\frac{a}{2} (\widehat{\bbox{z}} + \widehat{\bbox{x}})
\end{equation}
where $\widehat{\bbox{x}}$,
$\!$$\widehat{\bbox{y}}$,$\widehat{\bbox{z}}$
are unit vectors along the Cartesian axes.
A general lattice vector is
\begin{equation}\label{aeq2}
\bbox{R} = u\bbox{a}+v\bbox{b}+w\bbox{c}
~~;~~u,v,w~\mbox{are integers.}
\end{equation}
Similarly, the primitive translation vectors of the (bcc)
reciprocal lattice are :
\begin{eqnarray}\label{aeq3}
\bbox{A} &=& \frac{2\pi}{a}
(  \widehat{\bbox{x}} + \widehat{\bbox{y}}
                      - \widehat{\bbox{z}})~~~;~~~
\bbox{B} = \frac{2\pi}{a}
(- \widehat{\bbox{x}} + \widehat{\bbox{y}}
                      + \widehat{\bbox{z}})~~~; \nonumber   \\
\bbox{C} &=& \frac{2\pi}{a}
(  \widehat{\bbox{x}} - \widehat{\bbox{y}}
                      + \widehat{\bbox{z}}),    \nonumber
\end{eqnarray}
and the general reciprocal lattice vector is
\begin{equation}\label{aeq4}
\bbox{G}= h\bbox{A}+k\bbox{B}+l\bbox{C}
~~;~~h,k,l~\mbox{are integers.}
\end{equation}
Because we are interested specifically in projecting onto the
(110) plane, it is convenient to work with new unit vectors
$\widehat{\bbox{s}}$ and $\widehat{\bbox{t}}$ which lie along and
perpendicular to [110]:
\begin{equation}\label{aeq5}
 \widehat{\bbox{s}}=\frac{1}{\sqrt{2}}
(\widehat{\bbox{x}} + \widehat{\bbox{y}}) ~~;~~
 \widehat{\bbox{t}}=\frac{1}{\sqrt{2}}
(\widehat{\bbox{x}} - \widehat{\bbox{y}}),
\end{equation}
and thus
\begin{equation}\label{aeq6}
\bbox{R} =
R_s \widehat{\bbox{s}}+
R_t \widehat{\bbox{t}}+
R_z \widehat{\bbox{z}}
\end{equation}
with
\begin{eqnarray}
R_s &=& \frac{a}{2\sqrt{2}}(2u+v+w) ~~~;~~~
\label{aeq7}
R_t = \frac{a}{2\sqrt{2}}(-v+w) ~~~; \nonumber \\
R_z &=& \frac{a}{2}(v+w),
\end{eqnarray}
while
\begin{equation}\label{aeq8}
\bbox{G}=
G_s \widehat{\bbox{s}}+
G_t \widehat{\bbox{t}}+
G_z \widehat{\bbox{z}}
\end{equation}
with
\begin{eqnarray}\label{aeq9}
G_s &=& \frac{2\pi}{a}\sqrt{2}h ~~~;~~~
G_t = \frac{2\pi}{a}\sqrt{2}(-k+l) ~~~;\nonumber \\
G_z &=& \frac{2\pi}{a}(-h+k+l).
\end{eqnarray}
Likewise, we have
\begin{eqnarray}\label{aeq10}
\bbox{p}=
p_s \widehat{\bbox{s}}+
p_t \widehat{\bbox{t}}+
p_z \widehat{\bbox{z}}
~~\mbox{and}~~
\bbox{r}=
r_s \widehat{\bbox{s}}+
r_t \widehat{\bbox{t}}+
r_z \widehat{\bbox{z}}. \nonumber
\end{eqnarray}
The momentum density projected onto the (110) plane is
\begin{equation}\label{aeq11}
  \int \rho(\bbox{p}) dp_s = \rho(p_t,p_z).
\end{equation}
Having established the relevant notation, we write the momentum density
Eq. (\ref{eq2}) as
\begin{equation}\label{eq18}
\rho (\bbox{p}) = \rho (p_s,p_t,p_z)
 =  \sum_{b,\bbox{k}}
\int\!\!\!\int_{crystal} d\bbox{r}
                         d\bbox{r}'
           ~\psi_{b,\bbox{ k}}(\bbox{r})
\psi_{b,\bbox{k}}^{*}(\bbox{r}')
 e^{-i \bbox{p}\cdot(\bbox{r - r'})}.
\end{equation}
Writing the wavefunctions in the Bloch form and converting the
integral over $\bbox{r}$ in Eq. (\ref{eq2}) into a sum of
integrals over a single cell one obtains
\begin{eqnarray}
 \rho(p_t,p_z)&=&\int dp_s \sum_{b,\bbox{k}} \sum_{\bbox{R},\bbox{R'}}
        \int \int d\bbox{r} d\bbox{r'} u_{b,\bbox{k}} (\bbox{r})
        u_{b,\bbox{k}}^{*} (\bbox{r'})     \nonumber  \\
    && \times   e^{i\bbox{k} \cdot (\bbox{R}+\bbox{r})}
                e^{-i\bbox{k} \cdot (\bbox{R'}+\bbox{r'})}
                e^{-i\bbox{p} \cdot (\bbox{R}-\bbox{R'})}
                e^{-i\bbox{p} \cdot (\bbox{r}-\bbox{r'})}.
\label{eq12}
\end{eqnarray}
The two-dimensional projection of the occupation number density
$N(p_t,p_z)$ obtained by LCW-folding over {\em all} projected reciprocal
lattice vectors is
\begin{equation}\label{eq13}
    N(p_t,p_z) = \sum_{G_t,G_z} \rho(p_t+G_t,p_z+G_z)
\end{equation}
where the sum runs over all sets $(G_t,G_z)$.

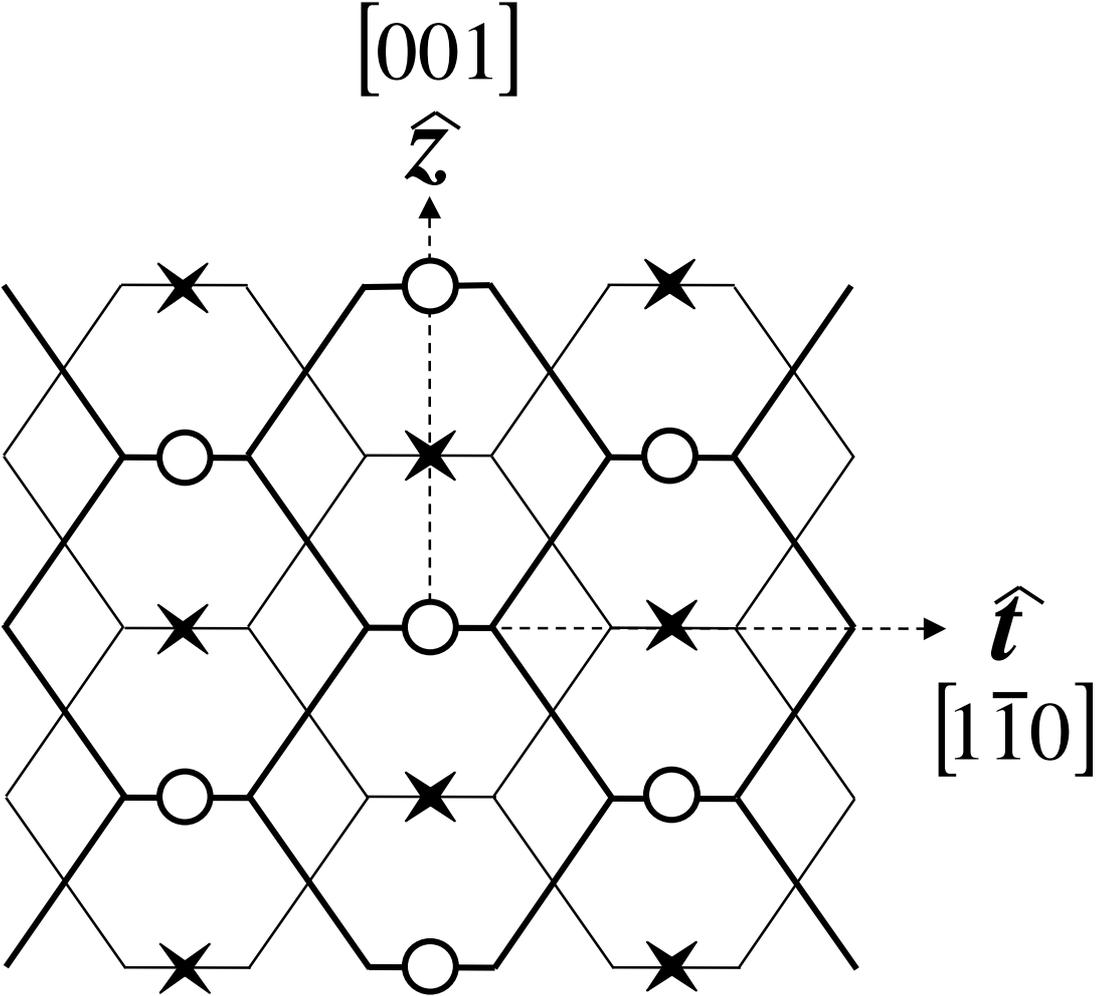
\begin{figure}
\caption{
The bcc reciprocal lattice points (corresponding to an fcc
crystal) are shown projected onto the (110) plane.
$\widehat{\bbox{z}}$ denotes the [001] and $\widehat{\bbox{t}}$
the [1$\bar{1}$0] direction. As discussed in the text, open
circles refer to points with even $h$ in Eq.\ (3.8) and crosses
to points with odd $h$. Outlines of the Brillouin zones for the
sublattices of circles and crosses are shown by thin and thick
solid lines respectively.
\label{fig1}}
\end{figure}

The formulation so far leading to Eq.\ (\ref{eq13}) has been
quite general. The nature of the problem as well as its solution
may be delineated with reference to Fig. \ref{fig1} which shows
schematically the projection of the bcc reciprocal lattice onto
the (110) plane. The lattice points in the (110) plane going
through the origin (for which $h=0$ in Eq. (3.8)) lie at the
positions of the circles. The two nearest (110) planes ($h = \pm
1$) lie at a distance of $G_s = (2\pi/a)\sqrt{2}$ in front of and
behind this central plane and the corresponding lattice points
project into the crosses. The next pair of planes projects to the
circles and then the process repeats itself. The entire bcc
reciprocal lattice is thus divided into two sublattices related
to the circles and the crosses with associated 2D Brillouin zones
shown by light and dark lines respectively.\cite{footnote2} It is
easily seen now that when the images of the Fermi surface given
by the periodic LCW-folded 3D-density are projected along [110],
the contributions from the circled and crossed sublattices will
overlap and get entangled. However, since the circled and crossed
sublattices are related by a simple shift of $2\pi/a$ along the
$z$ axis, if the contribution of one of these sublattices can be
suppressed in the folding procedure, the projection will no
longer suffer from the aforementioned complication.

The preceding strategy may be carried out more formally as
follows. In the $(s,t,z)$ space, Eqs.\ (\ref{aeq9}) give the
coordinates of various bcc lattice points generated via the
integers $h$, $k$ and $l$. Different layers perpendicular to the
[110] direction correspond to different values of $h$. It is
easily shown then that the circled sublattice in Fig.\
\ref{fig1} involves only {\em even} values of the index $h$
where $G_t$ and $G_z$ (ignoring prefactors in Eqs.\ (\ref{aeq9}))
are either both {\em even} or both {\em odd}. Similarly, the
crossed sublattice is given by {\em odd} values of $h$ where
$G_t$ and $G_z$ (again, ignoring prefactors) possess opposite
parities.

Bearing these considerations in mind, we divide the summation in
Eq.\ (\ref{eq13}) into two sums involving even and odd values of
$h$ and we use Eq.\ (\ref{eq12}). This yields
\begin{eqnarray}
    N(p_t,p_z) &=& \sum_{G_t,G_z}^{h:even} \rho(p_t+G_t,p_z+G_z)
               + \sum_{G_t,G_z}^{h:odd}
                                     \rho(p_t+G_t,p_z+G_z)   \nonumber  \\
               &=& \sum_{G_t,G_z}^{h:even}
                 \sum_{b,\bbox{k}} \sum_{\bbox{R},\bbox{R'}}
      \int dp_s \int \int d\bbox{r} d\bbox{r'} u_{b,\bbox{k}} (\bbox{r})
        u_{b,\bbox{k}}^{*} (\bbox{r'})
              e^{i\bbox{k} \cdot (\bbox{R}+\bbox{r})}
              e^{-i\bbox{k} \cdot (\bbox{R'}+\bbox{r'})}  \nonumber  \\
    && \times  e^{-i p_s (R_s - R'_s + r_s - r'_s)}
              e^{-i (p_t+G_t) (R_t - R'_t + r_t - r'_t)}
              e^{-i (p_z+G_z) (R_z - R'_z + r_z - r'_z)}  \nonumber  \\
    && +  \mbox{ a similar expression with $h$ odd}.
\label{eq14}
\end{eqnarray}
We now make use of the following identities:
\begin{eqnarray}
   \int dp_s e^{-ip_s(R_s-R'_s+r_s-r'_s)} & = &
             \delta_{R'_s R_s} \delta(r_s - r'_s)  \label{eq15} \\
   \sum_{G_t,G_z} e^{-iG_t(r_t-r'_t) -iG_z(r_z-r'_z)} & = &
       \delta(r_t - r'_t) \delta(r_z - r'_z)       \label{eq16} \\
   \sum_{R_t} \sum_{R_z} e^{i(k_t - p_t)R_t} e^{i(k_z - p_z)R_z} & = &
        \delta(k_t, p_t+G_t^a) \delta(k_z, p_z+G_z^a), \label{eq17}
\end{eqnarray}
where for the sake of simplicity we ignore forefactors, $\delta(a,b)$
denotes the Kronecker delta $\delta_{a,b}$, and $(G_t^a,G_z^a)$
are the components of an arbitrary reciprocal lattice vector
$\bbox{G^a}$.\cite{footnote3}
Furthermore,
\begin{equation}
\label{eq19}
        e^{-iG_t(R_t-R'_t) -iG_z(R_z-R'_z)}
   =    \left \{ \begin{array}{ll}
            1    &  \mbox{for even $h$}  \\
            e^{-\frac{2\pi}{a}i(R_z - R'_z)}  &  \mbox{for odd $h$}.
            \end{array}
            \right.
\end{equation}
The last identity is easily seen by using Eqs.\ (\ref{aeq7}) and
(\ref{aeq9}) and realizing that $v+w = (2/a) R_z$. The final
result is
\begin{equation}
N^{even}(p_t,p_z) = c \sum_{G^e_t,G^e_z}^{h:even}
\sum_{b,\bbox{k}}      \delta (k_t,p_t+G^e_t)\delta(k_z,p_z+G^e_z)
\label{aeq15}
\end{equation}
and
\begin{eqnarray}
N^{odd}(p_t,p_z)
 =  c\sum_{G^o_t,G^o_z}^{h:odd}\sum_{b,\bbox{k}}
                                  \delta (k_t,p_t+G^{o}_t)
\delta(k_z,p_z+\frac{2\pi}{a}+G^{o}_z),
\label{aeq16}
\end{eqnarray}
where the constant $c$ collects the irrelevant prefactors and
$G^{e(o)}_t$ and $G^{e(o)}_z$ are the components of
$\bbox{G}$ with even (odd) $h$.\cite{footnote3a}

The right hand side of Eq.\ (\ref{aeq15}) is just the projected
two-dimensional $\bbox{k}$ space occupation sampling function in
the repeated zone scheme for the set of reciprocal lattice
vectors with even $h$ in Eq.\ (\ref{aeq4}), while the right hand
side of Eq.\ (\ref{aeq16}) fulfills the same function for the set
with odd $h$. Thus, the set of Fermi surface images appearing in
Eq.\ (\ref{aeq15}), reappear in Eq.\ (\ref{aeq16}) but shifted by
$2\pi /a$ along $\widehat{\bbox{z}}$. By carrying out the LCW
folding over only one of these two sets of reciprocal lattice
vectors, the Fermi surfaces corresponding to the other set are
removed and thus a picture is obtained which is not complicated
by superposition.

Since $N^{even}$ and $N^{odd}$ are identical (apart from the shift)
and their sum yields the occupation function $N(p_t,p_z)$, it
follows that $N^{even}$ and $N^{odd}$ each carry half the intensity
of the occupation function.

Finally, it is clear that the present procedure can be used
straightforwardly in the analysis of positron annihilation 2D-ACAR
spectra from an fcc crystal.

%%%

\section{Experiment}
\label{sec:exp}
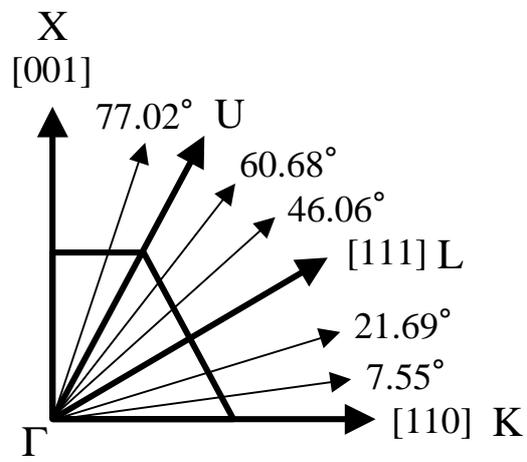
\begin{figure}
\caption{The nine directions in the (110) plane along which the
Compton profiles were measured and computed in this study.
\label{fig2}}
\end{figure}

A single crystal of Al-3at\%Li was grown from the melt by the
Bridgman-Stockbarger method under a pressurized argon atmosphere
to reduce loss of Li. The composition was determined by both
Atomic Absorption Spectrometry and Proton Induced Gamma Emission
and the final specimen, 10 mm in diameter and 1.5 mm thick, was
found to contain 3.0 at\% Li. The Compton profiles along nine
directions, all lying in the (110) plane (see Fig.\ \ref{fig2}),
were measured with the KEK-AR spectrometer.\cite{KEKAR} The
incident photon energy was 60 keV which is high enough so that
the impulse approximation would be satisfied quite well. The
overall momentum resolution in the measurements is estimated to
be 0.12 a.u. The total number of accumulated counts under the
profiles was about $1 \times 10^8$. The necessary energy dependent
corrections for absorption, detector and analyzer efficiency, and
scattering cross-section were applied. The contribution of double
scattering was simulated via the Monte Carlo program of
Sakai.\cite{26} The integrated intensity of the double scattering
events was found to be 10 \% of that of the single scattering
events. By using Eq.\ (\ref{eq4}), the nine measured profiles yield
the values of $B(z)$ along nine radial directions all of which lie
in the (110) plane in real space. $B(\bbox{r})$ was then
interpolated over an area of $50 \times 50$ a.u.$^2$ on a fine
square mesh of $500 \times 500$ $\bbox{r}$ points in the (110)
plane. The projection $\rho(p_t,p_z)$ of the 3D momentum density
was obtained via an inverse 2D-Fourier transformation of the
interpolated $B(\bbox{r})$. The $\rho(p_t,p_z)$ thus obtained was
partially folded following the procedure of Sec.\ \ref{sec:fold} to
obtain the 2D-occupation number density.

%%%

\section{Computations}\label{sec:the}
The computation of the electronic structure of Al-3at\%Li was
carried out within the all electron charge self-consistent
KKR-CPA framework and is parameter-free. A lattice constant of
$a=7.6534$ a.u. was assumed. The underlying KKR-CPA methodology
is described in Refs.\ \onlinecite{27,28,29,ref5,ref6}. The
relevant Green's function formulation for treating the momentum
density and Compton profile in disordered alloys is given in
Refs.\ \onlinecite{ref7} and \onlinecite{ref8}. Exchange and
correlation effects were incorporated within the von Barth-Hedin
local density approximation.\cite{lda,footnote3b} In order to obtain
the Compton profile the momentum density $\rho(\bbox{p})$ was first
evaluated over a fine mesh of $48 \times 4851 \times 1777$
$\bbox{p}$ points covering momenta up to $p_{\text{max}} \sim 5$
a.u. Compton profiles with scattering vectors along the same
directions as the experimental ones were then computed accurately
by evaluating the integral of Eq.\ (\ref{eq1}). The accuracy of
the computed profiles is about 1 part in 10$^4$. The nine
theoretical profiles were convoluted with the experimental
resolution and then treated in a manner identical to that of the
nine experimental profiles.

%%%

\section{Application to Al-3\%Li Alloy} \label{sec:appl}

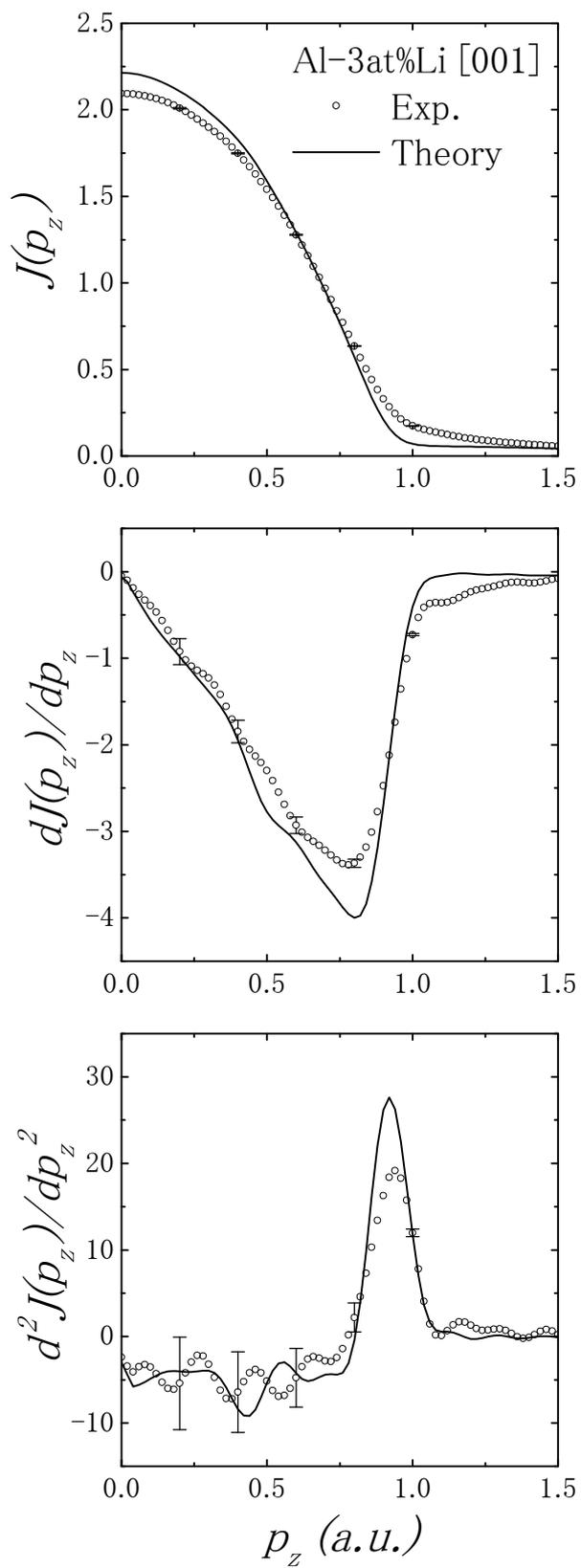
\begin{figure}
\caption{
Experimental and theoretical directional Compton profiles in
Al-3\%Li and the associated first and second derivatives are
compared along the [001] direction. The theoretical results have
been broadened to reflect the experimental resolution.
\label{fig3}}
\end{figure}

We briefly compare first the theoretical and experimental
directional profiles by taking the [100] CP as an example;
results along other directions are similar. After subtraction of
the core contribution the two profiles of Fig.\ \ref{fig3} are
area-normalized to yield the same number of electrons over the
range $\pm 4$ a.u. The theoretical CP is seen to be higher than
the experimental one at low momenta with the situation reversing
itself at high momenta. The Fermi cut-off in the first derivative
of the CP around 1 a.u. is sharper in the theory; this effect is
also evident in the second derivative where the corresponding
calculated peak is higher and narrower compared to the
measurements even though the theoretical CP's and their
derivatives include resolution broadening. Similar discrepancies
between theory and experiment have been observed previously in
other systems and can be ascribed to the failure of the present
LDA-based independent particle model to properly account for
correlations in the electron gas.\cite{5,6,7,8,9,10,11,12,13}
These correlation effects result in the shift of spectral weight
from below to above the Fermi momentum ($p_F$) and a reduction in
the size of the break ($Z_k$) in the momentum density at $p_F$.
Concerning fine structure, features around 0.3 a.u. and 0.6 a.u.
in the first derivative of the experimental CP are reproduced
reasonably by theory. These features are quite similar to those
found in Al in the positron annihilation (1D-angular correlation)
experiments\cite{okada} as well as in the recent high-resolution
Compton scattering measurements.\cite{ohata} In the case of Al,
these have been explained by the fact that the Fermi surface of
Al deviates from a sphere (in the extended zone) and bulges on
the hexagonal zone face near the W-K-W and W-U-W zone
edges;\cite{kubo} this mechanism is presumably at play in
Al-3\%Li as well since the addition of 3\% Li induces relatively
little change in the Fermi surface of Al.

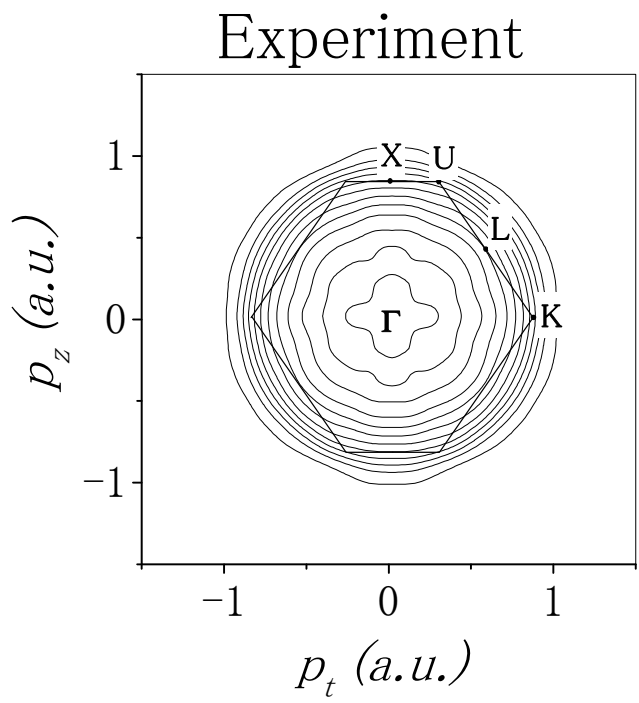
\begin{figure}
\caption{
Contour maps of the reconstructed experimental and theoretical
momentum densities in the (110) or $(p_z,p_t)$ plane. Note that
the results shown represent the 2D-distribution obtained after
the 3D momentum density is integrated along the [110] direction.
The first Brillouin zone boundary and the high symmetry points in
the (110) plane are indicated.
\label{fig4}}
\end{figure}

Before discussing the effects of folding, it is helpful to
consider Fig.\ \ref{fig4} which shows contour maps of the
reconstructed experimental and theoretical momentum density in
the (110) plane. We emphasize that an integration along [110]
(i.e., $p_s$) is implicit here as well as throughout this
article. Therefore, what is shown in all cases is the (110) 2D
projection of the underlying 3D momentum density distribution. As
expected, the data of Fig.\ \ref{fig4} display two-fold
crystalline symmetry, but it is difficult to ascertain the
details of the FS, except perhaps that it is roughly spherical in
shape (in the extended zone) with some bulging along the $\langle
110 \rangle$ directions.

Insight into the nature of our partial folding scheme is provided
by Fig.\ \ref{fig5} which considers the case of the free electron
model where the Fermi sphere corresponds to 2.94 electrons/atom
appropriate to the number of conduction electrons in Al-3\%Li.
Figure \ref{fig5}(a) has been obtained via the conventional
LCW-folding procedure in which {\em all} reciprocal lattice
vectors are used, whereas Fig.\ \ref{fig5}(b) includes the
contribution of only one of the sublattices as discussed in Sec.\
\ref{sec:fold}. The imprint of the free electron FS (repeated
periodically) is obvious in the 2D-occupation number density of
Fig.\ \ref{fig5}(b), even though this is far from being the case
in Fig.\ \ref{fig5}(a) which shows the appearance of four
cross-like features around the L symmetry points due to
overlapping Fermi surfaces. Furthermore, complicated structures
arise in Fig.\ \ref{fig5}(a) around the zone center. It is
striking how the two overlapping Fermi surfaces in Fig.\
\ref{fig5}(a) are disentangled in
Fig.\ \ref{fig5}(b).\cite{footnote4} These results show clearly
the power of the present partial folding approach for analyzing FS
images in terms of the (110)-projection in an fcc crystal.

The application to Al-3\%Li is considered in Fig.\ \ref{fig6}
which compares theoretical and experimental (110)-projected,
partially folded, 2D occupation densities. On the whole, the FS
contours in Fig.\ \ref{fig6} are rather like those in Fig.\
\ref{fig5}(b), indicating that the FS of the alloy is more or
less spherical, despite some deviations. Note that the FS
contours in Fig.\ \ref{fig6} are not symmetrical with respect to
the U-K lines, even though this is strictly true for the free
electron case in Fig.\ \ref{fig5}(b). This is a consequence of
the asphericity of the FS; the FS bulges more outside the
hexagonal zone face near the W-K-W zone edge (where two hexagonal
faces of the first zone intersect) than near the W-U-W zone edge
(intersection between a hexagonal and a square face of the first
zone). This can also be described as a slight flattening of the
FS in the vicinity of U, combined with a small expansion near K.
Theory and experiment are in good overall accord in Fig.\
\ref{fig6}, although the fact that the oblong maximum around the
L point in Fig.\ \ref{fig6}(b) is slightly more rotated compared
to Fig.\ \ref{fig6}(a) suggests that the aforementioned
asphericity of the FS is somewhat stronger in the experiment.

Finally, we consider radial cuts through the 2D momentum
distribution of Fig.\ \ref{fig4}; a typical cut (along
$\Gamma$-L) is shown in Fig.\ \ref{fig7}. It is well known that
the value of the Fermi momentum $p_F$ is not given correctly by
the inflection point in the profile of Fig.\ \ref{fig7}(a), or
equivalently, the position of the minimum in the first derivative
of Fig.\ \ref{fig7}(b).\cite{footnote5} The situation may be
simulated by considering the free electron case. Here the
2D-distribution corresponding to the profile of Fig.\
\ref{fig7}(a) is a semi-circle and the first derivative is
negative and diverges at $p_F$. When the semi-circle is
convoluted with the experimental resolution, a tail appears
beyond $p_{F}$ and the divergence of the first derivative at
$p_{F}$ disappears. The inflection point defining the minimum of
the first derivative then moves below $p_{F}$ to an extent which
depends on the width of the resolution function. Accordingly, by
using a free electron semi-circle of radius 0.919 a.u. and the
present experimental resolution of 0.12 a.u., we have determined
that the position of the inflection points in cuts such as those
of Fig.\ \ref{fig7} must be increased by 4.2 \% in order to
obtain the correct value of $p_F$. Applying this correction,
theoretical and experimental $p_F$ values in the (110) plane can
be mapped out from the data of Fig.\ \protect \ref{fig4}. The
results of the analysis are given in Fig.\ \ref{fig8}, and show
an excellent level of accord between theory and experiment.
Nevertheless, a careful examination of Fig.\ \ref{fig8} reveals a
slight contraction of the experimental FS (compared to theory)
near U in combination with a slight expansion near K, consistent
with our earlier discussion in connection with Fig.\ \ref{fig6}
above.

%%%

\section{Summary and Conclusions} \label{sec:sum}

We address the question of how Fermi surface (FS) signatures can
be enhanced in the 2D distribution obtained by projecting the 3D
momentum density along the [110] direction in an fcc crystal. The
standard LCW folding procedure invoked in this connection
involves a summation of the momentum density over all reciprocal
lattice vectors to obtain the electron occupation number density
in the system. This procedure has been used to produce a
reasonably clear FS imprint, even when the (110) projected
momentum density is considered, in several studies of bcc
crystals, but it has not been successful in fcc crystals. We
show, however, that the (110) projection of the LCW-folded
momentum density in the fcc lattice can be viewed as a
superposition of two equivalent FS images which are shifted by
$2\pi/a$ along the [001] axis, and that these two images can be
disentangled if one uses a selected subset of reciprocal lattice
vectors in the folding process. Our novel folding procedure is
particularly well suited for investigating fermiology-related
issues via high resolution Compton experiments where a full 3D
reconstruction of the data is far more demanding than a
2D-reconstruction using measured directional profiles all of
which involve scattering vectors in the (110) plane. The
technique will also be valuable in analyzing positron
annihilation (2D-ACAR) experiments in fcc lattices.

An extensive application of the aforementioned partial folding
procedure to the case of an fcc disordered alloy is presented. To
this end, we have measured nine high resolution directional
Compton profiles on a single crystal specimen of Al-3at\%Li with
scattering vectors in the (110) plane. We have also carried out
highly accurate computations of corresponding theoretical
profiles within the self-consistent, parameter free KKR-CPA
framework. Occupation number densities based on theoretical as
well as the experimental profiles obtained via partial folding
display clear images of the FS. A reasonable level of accord is
found between theory and experiment with respect to the overall
shape of the directional profiles as well as the fine structure
in the first and the second derivatives. The experimental results
with regard to the FS are also
in accord with KKR-CPA predictions, although there is
an indication of a slightly stronger bulging of the measured
Fermi surface outside the hexagonal zone face along the W-K-W
than along the W-U-W zone edge. Upon adding 3\% Li, the FS of Al
remains essentially free electron like, some overall shrinking of
its size due to the reduced e/a ratio notwithstanding. Large FS
sheets in Al thus show no sign of an anomaly which might
destabilize the $\alpha$ phase. A Fermi surface driven mechanism
in this connection will therefore need to focus on small features
in the FS of Al (electron pockets in the third conduction band)
which can undergo substantial change and may even disappear with
decreasing electron concentration or altered electron lattice
interaction upon alloying. \\

\acknowledgments
The Compton profile measurements were performed with the approval of
the Photon Factory Advisory Committee, Proposal number 97G288. This work 
is supported by the Polish Committee for Scientific Research, Grant 
number 2 P03B 028 14, the US Department of Energy contract
W-31-109-ENG-38, and benefited from the allocation of supercomputer
time at NERSC and the Northeastern University Advanced Scientific
Computation Center (ASCC).

%%%%

% CORRECTIONS TO FOLLOWING FIGURE--AB 12/19/00
%  +  signs are not marked in my version of the figure.

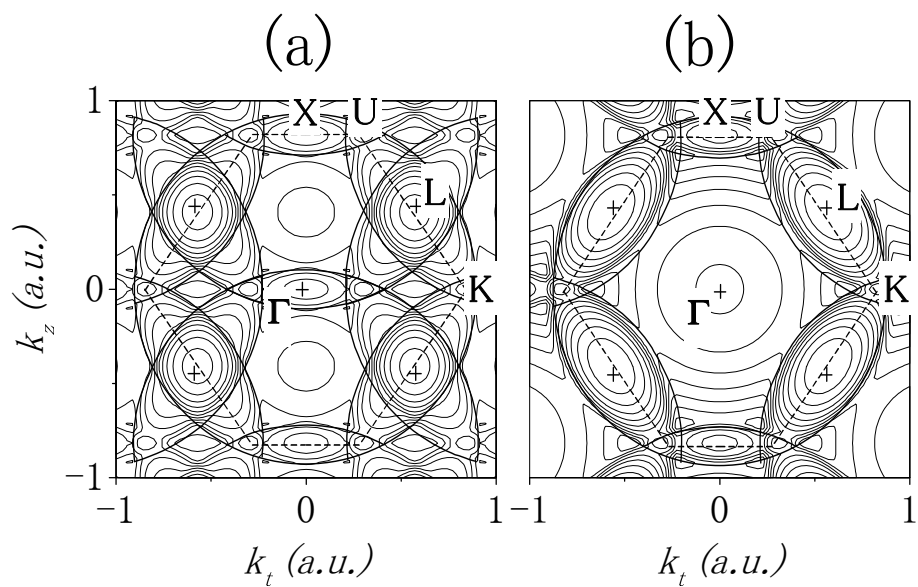
\begin{figure}
\caption{
Contour maps of the 2D occupation number density (after
projection along [110]), $N(k_z, k_t)$, in the (110) plane for
the free-electron model at the electron concentration of
Al-3\%Li. (a) Conventional LCW folding using all reciprocal
lattice vectors. (b) Present partial LCW folding scheme using
selected reciprocal lattice vectors. Experimental resolution
broadening is included in the calculations. The Brillouin zone
boundary and the projected free-electron sphere are shown. +
signs indicate high density regions.
\label{fig5}}
\end{figure}

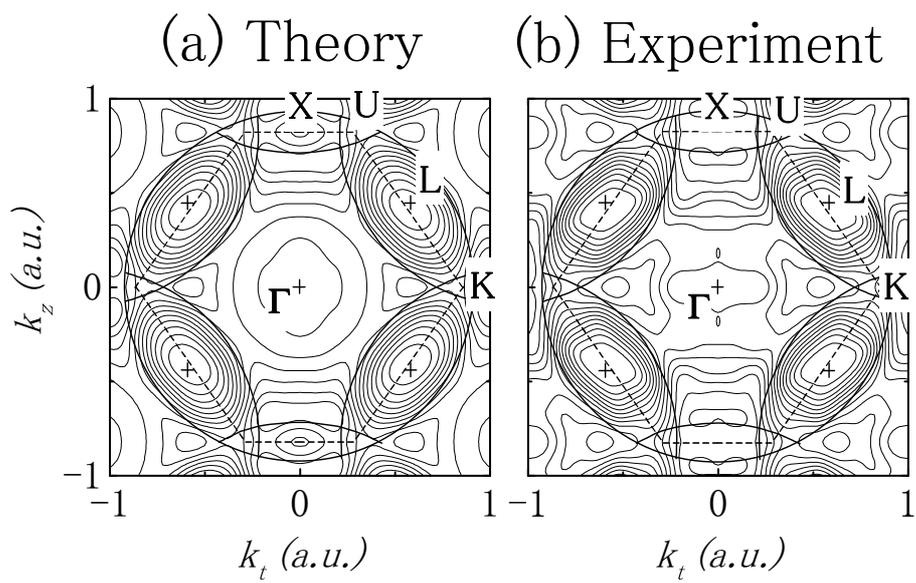
\begin{figure}
\caption{
Contour maps of the theoretical and experimental 2D occupation
number densities (after projection along [110]), $N(k_z, k_t)$,
in the (110) plane based on the present partial LCW-folding
scheme in Al-3\%Li. Resolution broadening is included in the
theory. See caption to Fig. \protect \ref{fig5} for other
notational details.
\label{fig6}}
\end{figure}

% CORRECTIONS TO FOLLOWING FIGURE--AB-12/19/00
%From PEM: Fig. 7 has \rho(p) along the axes, whereas that should be
%          \rho(p_t,p_z)! (this also applies to the latest figure)

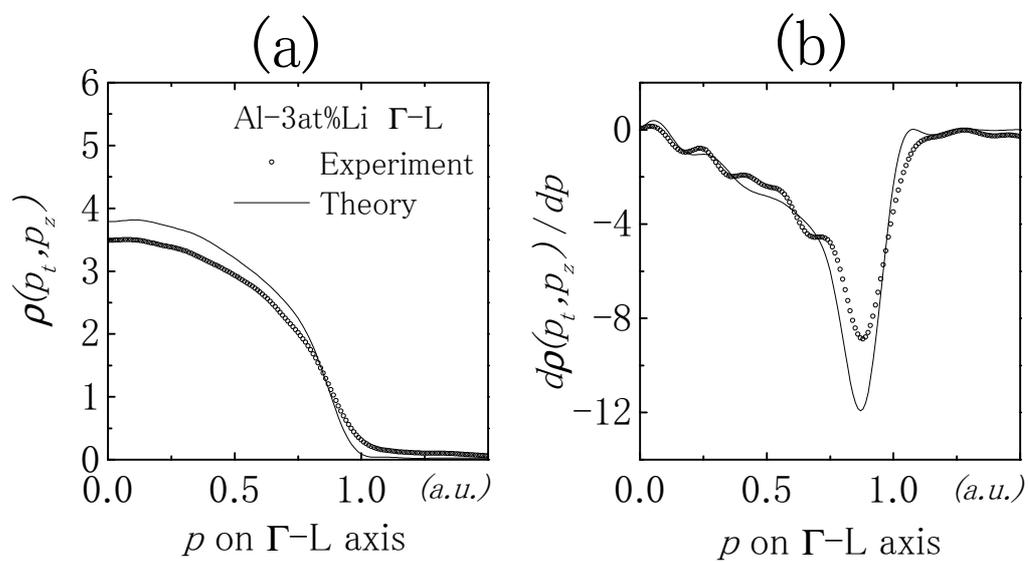
\begin{figure}
\caption{
Radial sections along the $\Gamma$-L direction through the
experimental and theoretical [110]-projected momentum
distributions of Fig. \protect \ref{fig4} are shown together with
the corresponding first derivatives.
\label{fig7}}
\end{figure}

% CORRECTIONS TO FOLLOWING FIGURE--AB-12/19/00
%  The BZ zone boundary lines are not connected around the U point-these
%  should be connected. The dashed circle for the Al case should be shown if
%  it looks reasonable, i.e. ressonably distinguished from the solid circle.
%  Make appropriate modifications in the caption below.

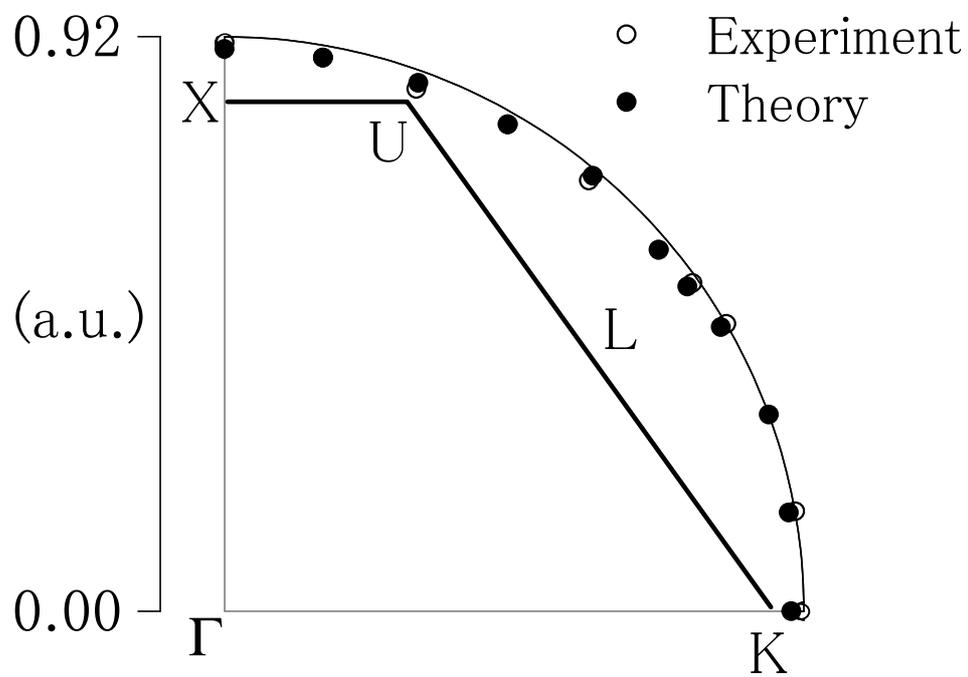
\begin{figure}
\caption{
Fermi momentum $p_F$ in the (110) plane in Al-3\%Li disordered
alloy. Open circles give results deduced from the present Compton
profile measurements, while the KKR-CPA based theoretical
predictions are given by filled circles. The solid circular arc
indicates the free electron momentum of 0.919 a.u. for Al-3\%Li.
\label{fig8}}
\end{figure}

%%%%


\begin{references}
\bibitem{footnote1}
  The form on the right hand side of Eq.\ (\ref{eq7}) is of course
  based on the independent particle model and thus assumes the absence
  of correlations inherent therein.
\bibitem{footnote1a}
  A high resolution Compton study of the [111] profile in Al-3\%Li is
  reported in Ref.\ \onlinecite{14}.
\bibitem{footnote2}
  This partitioning does not correspond to the more common one where
  the bcc lattice is viewed as a superposition of two simple cubic
  lattices shifted by half the cube diagonal with respect to each
  other.
\bibitem{footnote3}
  In evaluating Eq.\ (\ref{eq17}) it should be noted that the values of
  $R_t$ and $R_z$ in the summations are not independent but coupled by
  the symmetry of the direct lattice, expressed in Eq.\ (\ref{aeq1}).
  Thus, both $(k_t-p_t)R_t =2n \pi$ and $(k_t - p_t)R_t = (2n+1) \pi$
  lead to admissable solutions (namely $G_t,G_z$ both even and both
  odd, respectively, for even $h$) since in the second case the minus
  sign of the $z$ factor cancels that of the $t$ factor.
\bibitem{footnote3a}
  In a similar manner, by multiplying the left side of
  Eq.\ (\ref{eq19}) by $\exp[2\pi i(v-v')] = 1$ and noting that $-v+w =
  (2\sqrt{2}/a) R_t$, it can be shown that the two lattices differ by
  a shift of $2\pi\sqrt{2}/a$ along $\widehat{\bbox{t}}$.
\bibitem{footnote3b}
  Other, perhaps more satisfactory, forms of the exchange-correlation
  functional could of course be used in the process of obtaining the
  selfconsistent crystal potential. However, the effects are not
  particularly significant, especially within the context of this
  article.
\bibitem{footnote4}
  Of course, the remaining overlap in Fig.\ \ref{fig5}(b) is due to the
  size of the Fermi surfaces centered on neighboring $\Gamma$ points
  and is inherent in the electronic structure.
\end{references}
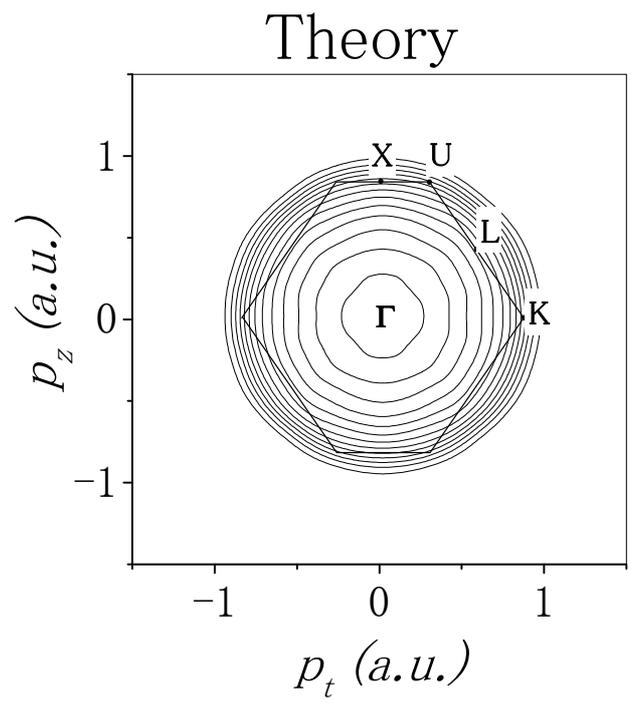
\end{document}